\documentstyle[twoside, epsfig]{article}

\catcode`\@=11
\long\def\@makefntext#1{
\protect\noindent \hbox to 3.2pt {\hskip-.9pt
$^{{\eightrm\@thefnmark}}$\hfil}#1\hfill}               

\def\@makefnmark{\hbox to 0pt{$^{\@thefnmark}$\hss}}    

\def\ps@myheadings{\let\@mkboth\@gobbletwo
\def\@oddhead{\hbox{}
\rightmark\hfil\eightrm\thepage}
\def\@oddfoot{}\def\@evenhead{\eightrm\thepage\hfil
\leftmark\hbox{}}\def\@evenfoot{}
\def\sectionmark##1{}\def\subsectionmark##1{}}

\oddsidemargin=\evensidemargin
\newcounter{sectionc}\newcounter{subsectionc}\newcounter{subsubsectionc}
\renewcommand{\section}[1] {\vspace{12pt}\addtocounter{sectionc}{1}
\setcounter{subsectionc}{0}\setcounter{subsubsectionc}{0}\noindent
        {\tenbf\thesectionc. #1}\par\vspace{5pt}}
\renewcommand{\subsection}[1] {\vspace{12pt}\addtocounter{subsectionc}{1}
      \setcounter{subsubsectionc}{0}\noindent
      {\bf\thesectionc.\thesubsectionc.{\kern1pt \bfit #1}}\par\vspace{5pt}}
\renewcommand{\subsubsection}[1]
      {\vspace{12pt}\addtocounter{subsubsectionc}{1}
      \noindent{\tenrm\thesectionc.\thesubsectionc.\thesubsubsectionc.
      {\kern1pt \tenit #1}}\par\vspace{5pt}}
\newcommand{\nonumsection}[1] {\vspace{12pt}\noindent{\tenbf #1}
        \par\vspace{5pt}}

\newcounter{appendixc}
\newcounter{subappendixc}[appendixc]
\newcounter{subsubappendixc}[subappendixc]
\renewcommand{\thesubappendixc}{\Alph{appendixc}.\arabic{subappendixc}}
\renewcommand{\thesubsubappendixc}
        {\Alph{appendixc}.\arabic{subappendixc}.\arabic{subsubappendixc}}

\renewcommand{\appendix}[1] {\vspace{12pt}
        \refstepcounter{appendixc}
        \setcounter{figure}{0}
        \setcounter{table}{0}
        \setcounter{lemma}{0}
        \setcounter{theorem}{0}
        \setcounter{corollary}{0}
        \setcounter{definition}{0}
        \setcounter{equation}{0}
        \renewcommand{\thefigure}{\Alph{appendixc}.\arabic{figure}}
        \renewcommand{\thetable}{\Alph{appendixc}.\arabic{table}}
        \renewcommand{\theappendixc}{\Alph{appendixc}}
        \renewcommand{\thelemma}{\Alph{appendixc}.\arabic{lemma}}
        \renewcommand{\thetheorem}{\Alph{appendixc}.\arabic{theorem}}
        \renewcommand{\thedefinition}{\Alph{appendixc}.\arabic{definition}}
        \renewcommand{\thecorollary}{\Alph{appendixc}.\arabic{corollary}}
        \renewcommand{\theequation}{\Alph{appendixc}.\arabic{equation}}
        \noindent{\tenbf Appendix \theappendixc #1}\par\vspace{5pt}}
\newcommand{\subappendix}[1] {\vspace{12pt}
        \refstepcounter{subappendixc}
        \noindent{\bf Appendix \thesubappendixc. {\kern1pt \bfit #1}}
        \par\vspace{5pt}}
\newcommand{\subsubappendix}[1] {\vspace{12pt}
        \refstepcounter{subsubappendixc}
        \noindent{\rm Appendix \thesubsubappendixc. {\kern1pt \tenit #1}}
        \par\vspace{5pt}}

\newcommand{\smalllineskip}{\baselineskip=10pt}

\def\eightcirc{
\begin{picture}(0,0)
\put(4.4,1.8){\circle{6.5}}
\end{picture}}
\def\eightcopyright{\eightcirc\kern2.7pt\hbox{\eightrm c}}

\def\abstracts#1#2#3{{
        \centering{\begin{minipage}{4.5in}\baselineskip=10pt\footnotesize
        \parindent=0pt #1\par
        \parindent=15pt #2\par
        \parindent=15pt #3
        \end{minipage}}\par}}

\renewenvironment{thebibliography}[1]
        {\frenchspacing
         \ninerm\baselineskip=11pt
         \begin{list}{\arabic{enumi}.}
        {\usecounter{enumi}\setlength{\parsep}{0pt}
         \setlength{\leftmargin 12.7pt}{\rightmargin 0pt} 
         \setlength{\itemsep}{0pt} \settowidth
        {\labelwidth}{#1.}\sloppy}}{\end{list}}

\newcounter{itemlistc}
\newcounter{romanlistc}
\newcounter{alphlistc}
\newcounter{arabiclistc}

\newcommand{\fcaption}[1]{
        \refstepcounter{figure}
        \setbox\@tempboxa = \hbox{\footnotesize Fig.~\thefigure. #1}
        \ifdim \wd\@tempboxa > 5in
           {\begin{center}
        \parbox{5in}{\footnotesize\smalllineskip Fig.~\thefigure. #1}
            \end{center}}
        \else
             {\begin{center}
             {\footnotesize Fig.~\thefigure. #1}
              \end{center}}
        \fi}

\newcommand{\tcaption}[1]{
        \refstepcounter{table}
        \setbox\@tempboxa = \hbox{\footnotesize Table~\thetable. #1}
        \ifdim \wd\@tempboxa > 5in
           {\begin{center}
        \parbox{5in}{\footnotesize\smalllineskip Table~\thetable. #1}
            \end{center}}
        \else
             {\begin{center}
             {\footnotesize Table~\thetable. #1}
              \end{center}}
        \fi}

\def\@citex[#1]#2{\if@filesw\immediate\write\@auxout
        {\string\citation{#2}}\fi
\def\@citea{}\@cite{\@for\@citeb:=#2\do
        {\@citea\def\@citea{,}\@ifundefined
        {b@\@citeb}{{\bf ?}\@warning
        {Citation `\@citeb' on page \thepage \space undefined}}
        {\csname b@\@citeb\endcsname}}}{#1}}

\newif\if@cghi
\def\cite{\@cghitrue\@ifnextchar [{\@tempswatrue
        \@citex}{\@tempswafalse\@citex[]}}
\def\citelow{\@cghifalse\@ifnextchar [{\@tempswatrue
        \@citex}{\@tempswafalse\@citex[]}}
\def\@cite#1#2{{$\null^{#1}$\if@tempswa\typeout
        {IJCGA warning: optional citation argument
        ignored: `#2'} \fi}}

\def\@refcitex[#1]#2{\if@filesw\immediate\write\@auxout
        {\string\citation{#2}}\fi
\def\@citea{}\@refcite{\@for\@citeb:=#2\do
        {\@citea\def\@citea{, }\@ifundefined
        {b@\@citeb}{{\bf ?}\@warning
        {Citation `\@citeb' on page \thepage \space undefined}}
        \hbox{\csname b@\@citeb\endcsname}}}{#1}}

\def\@refcite#1#2{{#1\if@tempswa\typeout
        {IJCGA warning: optional citation argument
        ignored: `#2'} \fi}}

\def\refcite{\@ifnextchar[{\@tempswatrue
        \@refcitex}{\@tempswafalse\@refcitex[]}}

\def\pmb#1{\setbox0=\hbox{#1}
        \kern-.025em\copy0\kern-\wd0
        \kern.05em\copy0\kern-\wd0
        \kern-.025em\raise.0433em\box0}

\def\fnt#1#2{\footnotetext{\kern-.3em
        {$^{\mbox{\scriptsize #1}}$}{#2}}}

\def\fpage#1{\begingroup
\voffset=.3in
\thispagestyle{empty}\begin{table}[b]\centerline{\footnotesize #1}
        \end{table}\endgroup}

\def\runninghead#1#2{\pagestyle{myheadings}
\markboth{{\protect\footnotesize\it{\quad #1}}\hfill}
{\hfill{\protect\footnotesize\it{#2\quad}}}}
\font\tenrm=cmr10
\font\tenit=cmti10
\font\tenbf=cmbx10
\font\bfit=cmbxti10 at 10pt
\font\ninerm=cmr9

\font\eightrm=cmr8

\def\qed{\hbox{${\vcenter{\vbox{                      
   \hrule height 0.4pt\hbox{\vrule width 0.4pt height 6pt
   \kern5pt\vrule width 0.4pt}\hrule height 0.4pt}}}$}}

\begin{document}

\runninghead {Rainer W. K\"uhne}
{Possible observation of $\ldots$}

\thispagestyle{empty}\setcounter{page}{36}
\vspace*{0.88truein}
\fpage{36}

\centerline{\bf POSSIBLE OBSERVATION OF A SECOND KIND OF LIGHT}
\vspace*{0.035truein}
\centerline{\bf  -- MAGNETIC PHOTON RAYS}
\vspace*{0.035truein}

\vspace*{0.37truein}
\centerline{\footnotesize Rainer W. K\"uhne}

\centerline{\footnotesize \it
Lechstr. 63, 38120 Braunschweig, Germany}
\baselineskip=10pt
\centerline{\footnotesize \it
kuehne70@gmx.de}


\baselineskip 5mm

\vspace*{0.21truein}

\abstracts{Several years ago, I suggested a quantum field theory which has many  attractive features. (1) It can explain the quantization of electric charge. (2) It describes symmetrized Maxwell equations. (3) It is manifestly covariant. (4) It describes local four-potentials. (5) It avoids the  unphysical Dirac string. My model predicts a second kind of light, which I named ``magnetic photon rays.''  
Here I will discuss possible observations of this 
radiation by August Kundt in 1885, Alipasha Vaziri in February 2002, and 
Roderic Lakes in June 2002.}{}{}


\bigskip

$$$$

\section{The Theoretical Background}

\subsection{The Model}

The existence of the second kind of light was predicted theoretically. 
It can be understood by the following argumentation.

In 1948/1949 Tomonaga, Schwinger, Feynman, and Dyson introduced quantum 
electrodynamics [1]. It is the quantum field theory of electric and 
magnetic phenomena. This theory has one shortcoming. 
It cannot explain why electric charge is quantized, i.e. why it 
appears only in discrete units.

In 1931 Dirac [2] introduced the concept of magnetic monopoles. 
He has shown that any theory which includes magnetic monopoles 
requires the quantization of electric charge.

A theory of electric and magnetic phenomena which includes Dirac 
monopoles can be formulated in a manifestly covariant and 
symmetrical way if two four-potentials are used. Cabibbo and Ferrari 
in 1962 [3] were the first to formulate such a theory. It was 
examined in greater detail by later authors [4 -- 6]. 
Within the 
framework of a quantum field theory one four-potential corresponds 
to Einstein's electric photon from 1905 [7]  
and the other four-potential 
corresponds to Salam's magnetic photon from 1966 [5].

In 1997 I have shown that the Lorentz force between an electric 
charge and a magnetic charge can be generated as follows [6]. 
An electric charge 
couples via the well-known vector coupling with an electric photon and 
via a new type of tensor coupling, named velocity coupling, with a 
magnetic photon. This velocity coupling requires the existence of a 
velocity operator.

For scattering processes this velocity is the relative velocity 
between the electric charge and the magnetic charge just before 
the scattering. For emission and absorption processes there is no 
possibility of a relative velocity. The velocity is the absolute 
velocity of the electric charge just before the reaction.

The absolute velocity of a terrestrial laboratory was measured by 
the dipole anisotropy of the cosmic microwave background radiation. 
This radiation was detected in 1965 by Penzias and Wilson [8], its 
dipole anisotropy was detected in 1977 by Smoot, Gorenstein, and Muller 
[9]. 
The mean value of the laboratory's absolute velocity is 371 km/s. 
It has an annual sinusoidal period because of the Earth's motion 
around the Sun with 30 km/s. It has also a diurnal sinusoidal period 
because of the Earth's rotation with 0.5 km/s.

According to my model from 1997 [6] each process that produces 
electric photons does create also magnetic photons. The cross-section 
of magnetic photons in a terrestrial laboratory is roughly one 
million times smaller than that of electric photons of the same energy. 
The exact value varies with time and has both the annual and the 
daily period.

As a consequence, magnetic photons are one million times harder to 
create, to shield, and to absorb than electric photons of the same 
energy.

The electric-magnetic duality is: 

\begin{center}
\begin{tabular}{lll}
electric charge & --- & magnetic charge \\
electric current & --- & magnetic current \\
electric conductivity & --- & magnetic conductivity \\
electric field strength & --- & magnetic field strength \\
electric four-potential & --- & magnetic four-potential \\
electric photon & --- & magnetic photon \\
electric field constant & --- & magnetic field constant \\
dielectricity number & --- & magnetic permeability
\end{tabular}
\end{center}

The refractive index of an insulator is the square root of the product of 
the dielectricity number and the magnetic permeability. Therefore it  
is invariant under a dual transformation. This means that electric and 
magnetic photon rays are reflected and refracted by insulators in the same 
way. Optical lenses cannot distinguish between electric and 
magnetic photon rays.

By contrast, electric and magnetic photon rays are reflected and refracted 
in a different way by metals. This is because electric conductivity and 
magnetic conductivity determine the reflection of light and they 
are not identical. The electric conductivity of a metal is several 
orders larger than the magnetic conductivity. 

\subsection{The Formulae for Classical Electromagnetodynamics}
Let $J^{\mu}=(P, {\bf J})$ denote the electric four-current and 
$j^{\mu}=(\rho , {\bf j})$ the magnetic four-current. The 
well-known four-potential of the electric photon is 
$A^{\mu}=(\Phi , {\bf A})$.  The four-potential of the magnetic photon is 
$a^{\mu}=(\varphi , {\bf a})$. Expressed in three-vectors the symmetrized 
Maxwell equations read,
\begin{eqnarray}
\nabla\cdot {\bf E} & = & P \\
\nabla\cdot {\bf B} & = & \rho \\
\nabla\times {\bf E} & = & - {\bf j} - \partial_{t} {\bf B} \\
\nabla\times {\bf B} & = & + {\bf J} + \partial_{t} {\bf E}
\end{eqnarray}
and the relations between field strengths and potentials are
\begin{eqnarray}
{\bf E} & = & - \nabla\Phi - \partial_{t} {\bf A} -\nabla\times {\bf a} \\
{\bf B} & = & - \nabla\varphi - \partial_{t} {\bf a} +\nabla\times {\bf A}.
\end{eqnarray}
By using the tensors
\begin{eqnarray}
F^{\mu\nu} & \equiv & \partial^{\mu}A^{\nu}- \partial^{\nu}A^{\mu} \\
f^{\mu\nu} & \equiv & \partial^{\mu}a^{\nu}- \partial^{\nu}a^{\mu}
\end{eqnarray}
we obtain the two Maxwell equations
\begin{eqnarray}
J^{\mu} & = & \partial_{\nu}F^{\nu\mu} = \partial^{2}A^{\mu} 
- \partial^{\mu}\partial^{\nu}A_{\nu} \\
j^{\mu} & = & \partial_{\nu}f^{\nu\mu} = \partial^{2}a^{\mu} 
- \partial^{\mu}\partial^{\nu}a_{\nu}.
\end{eqnarray}
Evidently, the two Maxwell equations are invariant under the 
$U(1)\times U'(1)$ gauge transformations
\begin{eqnarray}
A^{\mu} & \rightarrow & A^{\mu}-\partial^{\mu}\Lambda \\
a^{\mu} & \rightarrow & a^{\mu}-\partial^{\mu}\lambda .
\end{eqnarray}
Furthermore, the four-currents satisfy the continuity equations
\begin{equation}
0=\partial_{\mu}J^{\mu}= \partial_{\mu}j^{\mu}.
\end{equation}
The electric and magnetic field are related to the tensors above by
\begin{eqnarray}
E^{i} & = & F^{i0}- \frac{1}{2}\varepsilon^{ijk}f_{jk} \\
B^{i} & = & f^{i0}+ \frac{1}{2}\varepsilon^{ijk}F_{jk}.
\end{eqnarray}
Finally, the Lorentz force is
\begin{equation}
K^{\mu}  = Q(F^{\mu\nu}+ \frac{1}{2}\varepsilon^{\mu\nu\varrho\sigma}
              f_{\varrho\sigma})u_{\nu} 
 + q(f^{\mu\nu}- \frac{1}{2}\varepsilon^{\mu\nu\varrho\sigma}
              F_{\varrho\sigma})u_{\nu},
\end{equation}
where $\varepsilon^{\mu\nu\varrho\sigma}$ denotes the totally 
antisymmetric tensor. 

\section{Arguments for an Absolute Rest Frame}

\noindent
Soon after I presented my model of magnetic monopoles [6], I 
learned that the main obstacle for most physicists to accept my 
model was that it requires an absolute rest frame. For this reason, 
I will present the arguments for an absolute frame in this  
section. The first subsection deals with the classical arguments, 
the second subsection deals with the arguments based on General Relativity 
and relativistic cosmology.

\subsection{Space and Time Before General Relativity}

According to Aristotle, the Earth was resting in the centre of the universe. 
He considered the terrestrial frame as a preferred frame and all motion 
relative to the Earth as absolute motion. Space and time were absolute 
[10].

In the days of Galileo the heliocentric model of Copernicus 
[11] was valid. The Sun was thought to be resting within the 
centre of the universe and defining a preferred frame. Galileo argued that 
only relative motion was observed but not absolute motion. However, to 
fix motion he considered it as necessary to have not only relative motion, 
but also absolute motion [12].

Newton introduced the mathematical description of Galileo's kinematics. 
His equations described only relative motion. Absolute motion did not 
appear in his equations [13].

This inspired Leibniz to suggest that absolute motion is not required 
by the classical mechanics introduced by Galileo and Newton [14]. 

Huyghens introduced the wave theory of light. According to his theory, 
light waves propagate via oscillations of a new medium which consists 
of very tiny particles, which he named aether particles. He considered 
the rest frame of the luminiferous aether as a preferred frame 
[15].

The aether concept reappeared in Maxwell's theory of classical 
electrodynamics [16]. Faraday [17] unified Coulomb's 
theory of electricity [18] with Amp\`ere's theory of magnetism 
[19]. Maxwell unified Faraday's 
theory with Huyghens' wave theory of light, where in Maxwell's theory 
light is considered as an oscillating electromagnetic wave which 
propagates through the luminiferous aether of Huyghens.

We all know that the classical kinematics was replaced by Einstein's 
Special Relativity [20]. Less known is that Special Relativity is not 
able to answer several problems that were explained by classical mechanics.

According to the relativity principle of Special Relativity, all inertial 
frames are equivalent, there is no preferred frame. Absolute motion is not 
required, only the relative motion between the inertial frames is needed. 
The postulated absence of an absolute frame prohibits the existence of 
an aether [20].

According to Special Relativity, each inertial frame has its own relative 
time. One can infer via the 
Lorentz transformations [21] on the time of the other inertial 
frames. Absolute space and time do not exist. Furthermore, space is 
homogeneous and isotropic, there does not exist any rotational axis of 
the universe.

It is often believed that the Michelson-Morley experiment [22] 
confirmed the relativity principle and refuted the existence of a 
preferred frame. This believe is not correct. In fact, the result of 
the Michelson-Morley experiment disproved the existence of a preferred 
frame only if Galilei invariance is assumed. The experiment can be 
completely explained by using Lorentz invariance alone, the relativity 
principle is not required.

By the way, the relativity principle is not a phenomenon that belongs 
solely to Special Relativity. According to Leibniz it can be applied also 
to classical mechanics.

Einstein's theory of Special Relativity has three problems.

(i) The space of Special Relativity is empty. There are no entities apart 
from the observers and the observed objects in the inertial frames. 
By contrast, the space of classical mechanics can be filled with, say, 
radiation or turbulent fluids.

(ii) Without the concept of an aether Special Relativity can only 
describe but not explain why electric and magnetic fields oscillate in 
propagating light waves.

(iii) Special Relativity does not satisfy the equivalence principle 
[23] of General Relativity, according to which inertial mass and 
gravitational mass are identical. Special Relativity considers only 
inertial mass.

Special Relativity is a valid approximation of reality which is appropriate 
for the description of most of the physical phenomena examined until 
the beginning of the twenty-first century. However, the macroscopic 
properties of space and time are better described by General Relativity.

\subsection{General Relativity: Absolute Space and Time}

\noindent
In 1915 Einstein presented the field equations of General Relativity 
[24] and in 1916 he presented the first comprehensive article on 
his theory [25]. In a later work he showed an analogy between 
Maxwell's theory and General Relativity. The solutions of the free 
Maxwell equations are electromagnetic waves while the solutions of the 
free Einstein field equations are gravitational waves which propagate 
on an oscillating metric [26]. As a consequence, Einstein 
called space the aether of General Relativity [27]. However, 
even within the framework of General Relativity do electromagnetic waves 
not propagate through a luminiferous aether.

Einstein applied the field equations of General Relativity on the entire 
universe [28]. He presented a solution of a homogeneous, 
isotropic, and static universe, where the space has a positive 
curvature. This model became known as the Einstein universe. However, 
de Sitter has shown that the Einstein universe is not stable against 
density fluctuations [29].

This problem was solved by Friedmann and Lema\^itre who suggested a 
homogeneous and isotropic expanding universe where the space is curved 
[30].

Robertson and Walker presented a metric for a homogeneous and isotropic 
universe [31]. According to G\"odel this metric requires an 
absolute time [32]. In any homogeneous and isotropic cosmology 
the Hubble constant [33] and its inverse, the Hubble age of 
the universe, are absolute and not relative quantities. In the 
Friedmann-Lema\^itre universe there exists a relation between the actual 
age of the universe and the Hubble age.

According to Bondi and Gold, a preferred motion is given at each point 
of space by cosmological observations, namely the redshift-distance 
relation generated by the Hubble effect. It appears isotropic only 
for a unique rest frame [34].

I argued that the Friedmann-Lema\^itre universe has a finite age and 
therefore a finite light cone. The centre-of-mass frame of this Hubble 
sphere can be regarded as a preferred frame [6].

After the discovery of the cosmic microwave background radiation by 
Penzias and Wilson [8], it was predicted that it should have 
a dipole anisotropy generated by the Doppler effect by the Earth's 
motion. This dipole anisotropy was predicted in accordance with 
Lorentz invariance [35] and later discovered experimentally 
[9]. Peebles called these experiments ``aether drift 
experiments'' [36].

The preferred frames defined by the Robertson-Walker metric, the 
Hubble effect, and the cosmic microwave background radiation are 
probably identical. In this case the absolute motion of the Sun was 
determined by the dipole anisotropy experiments of the 
cosmic microwave background radiation to be $(371 \pm 1)$ km/s.

\section{Three Experiments to Verify the Magnetic Photon Rays}
\subsection{How to Verify the Magnetic Photon Rays}

\noindent
The easiest test to verify/falsify the magnetic photon is to illuminate a 
metal foil of thickness $1,\ldots ,100\mu$m  by a laser beam (or any other 
bright light source) and to place a detector (avalanche diode or 
photomultiplier tube) behind the foil. If a single foil is used, then the 
expected reflection losses are less than 1\%. If a laser beam of the 
visible light is used, then the absorption losses are less than 15\%. My 
model [6] predicts the detected intensity of the radiation to be 
\begin{equation}
f = r(v/c)^4
\end{equation}
times the intensity that would be detected 
if the metal foil were removed and the laser beam would directly illuminate 
the detector. Here
\begin{equation}
v = v_{sun} + v_{earth}\cos (2\pi t/T_e ) \cos ( \varphi_{ec}) 
+ v_{rotation} \cos(2\pi t/T_{rot}) \cos ( \varphi_{eq})
\end{equation}
is the absolute velocity of the laboratory. The absolute velocity 
of the Sun as measured by the dipole anisotropy of the cosmic microwave 
background radiation is
\begin{equation}
v_{sun} = (371 \pm 0.5) \mbox{km/s}.
\end{equation}
The mean velocity of the Earth around the Sun is
\begin{equation}
v_{earth} = 30 \mbox{km/s}.
\end{equation}
The rotation velocity of the Earth is
\begin{equation}
v_{rotation} = 0.5 \mbox{km/s} \cos ( \varphi ).
\end{equation}
The latitude of the dipole with respect to the ecliptic is
\begin{equation}
\varphi_{ec} = 15^{\circ}.
\end{equation}
The latitude of the dipole with respect to the equator (declination) is
\begin{equation}
\varphi_{eq} = 7^{\circ}.
\end{equation}
The latitude of the laboratory is
\begin{equation}
\varphi = 48^{\circ}
\end{equation}
for Strassbourg and Vienna and $\varphi = 43^{\circ}$ for Madison. 
The sidereal year is
\begin{equation}
T_e = 365.24 \mbox{days}.
\end{equation}
A sidereal day is
\begin{equation}
T_{rot} = 23\mbox{h}~ 56\mbox{min}.
\end{equation}
The zero point of the time, $t = 0$, is reached on December 9 at 0:00 local 
time. The speed of light is denoted by $c$. The factor for losses by 
reflection and absorption of magnetic photon rays of the visible light 
for a metal foil of thickness $1, \ldots ,100 \mu$m is
\begin{equation}
r = 0.8, \ldots , 1.0 .
\end{equation}
To conclude, my model [6] predicts the value $f\sim 10^{-12}$. 

More precisely, this value is correct only for interactions of 
free electric charges with photons. In these situations the cross-section 
of magnetic photons is reduced by the factor $(v/c)^{2}$ for emission 
and absorption processes with respect to the cross-section of 
magnetic photons of the same energy. Since in metals we do not have 
free electric charges nor free photons, this value has to be modified.

\subsection{The Experiment by August Kundt}

\noindent
In Strassbourg in 1885, August Kundt [37] passed sunlight through 
red glass, a polarizing 
Nicol, and platinized glass which was covered by an iron layer. The entire 
experimental setup was placed within a magnetic field. With the naked eye, 
Kundt measured the Faraday rotation of the polarization plane generated by 
the transmission of the sunlight through the iron layer. His result was a 
constant maximum rotation of the polarization plane per length of 
$418,000^{\circ}$/cm or $1^{\circ}$ per 23.9nm. He verified this result 
until thicknesses of up to 210nm and rotations of up to $9^{\circ}$. 

In one case, on a very clear day, he observed the penetrating sunlight for 
rotations of up to $12^{\circ}$. Unfortunately, he has not given the 
thickness of this particular iron layer he used. But if his result of a 
constant maximum rotation per length can be applied, then the corresponding 
layer thickness was $\sim 290$nm.

Let us recapitulate some classical electrodynamics to determine the 
behavior of light within iron. The penetration depth of light in a 
conductor is
$$
\delta = \frac{\lambda}{2\pi\gamma},
$$
where the wavelength in vacuum can be expressed by its frequency 
according to $\lambda = 1/ \sqrt{\nu^2 \varepsilon_0 \mu_0}$. The 
extinction coefficient is
$$
\gamma = \frac{n}{\sqrt{2}}\left[ -1 + \left( 1+ \left( 
\frac{\sigma}{2\pi\nu\varepsilon_0\varepsilon_r} \right)^2 \right) ^{1/2} 
\right] ^{1/2} ,
$$
where the refractive index is $n=\sqrt{\varepsilon_r \mu_r }$. For 
metals we get the very good approximation
$$
\delta\approx\left( \frac{1}{\pi\mu_0\mu_r\sigma\nu} \right) ^{1/2}.
$$
The specific resistance of iron is
$$
1/ \sigma = 8.7\times 10^{-8}\Omega\mbox{m},
$$
its permeability is $\mu_r \geq 1$. For red light of $\lambda =630$nm 
and $\nu =4.8\times 10^{14}$Hz we get the penetration depth
$$
\delta = 6.9\mbox{nm}.
$$

Only a small fraction of the sunlight can enter the iron layer. Three 
effects have to be considered. (i) The red glass allows the penetration 
of about $\varepsilon_1 \sim 50\% $ of the sunlight only. (ii) Only 
$\varepsilon_2 =2/ \pi \simeq 64\% $ of 
the sunlight can penetrate the polarization filter. (iii) Reflection 
losses at the surface of the iron layer have to be considered. The 
refractive index for electric photon light is given by

\begin{equation}
\bar n^{2} = \frac{n^{2}}{2} \left( 1+ \sqrt{ 1+ \left( 
\frac{\sigma}{2\pi\varepsilon_0 \varepsilon_r \nu} \right)^{2}} \right).
\end{equation}

For metals we get the very good approximation

\begin{equation}
\bar n \simeq \sqrt{ \frac{\mu_r \sigma}{4\pi\varepsilon_0 \nu}}.
\end{equation}
The fraction of the sunlight which is not reflected is
\begin{equation}
\varepsilon_3 = \frac{2}{1+ \bar n}= 
\frac{2}{1+ \sqrt{\mu_r \sigma /(4\pi\varepsilon_0 \nu )}}
\end{equation}
and therefore $\varepsilon_3 \simeq 0.13$ for the system considered. Taken 
together, 
the three effects allow only 
$\varepsilon_1 \varepsilon_2 \varepsilon_3 \sim 4\% $ 
of the sunlight to enter the iron layer. 

The detection limit of the naked eye is $10^{-13}$ times the brightness 
of sunlight provided the light source is pointlike. For an extended 
source the detection limit depends on the integral and the surface 
brightness. The detection limit for a source as extended as the Sun 
(0.5$^{\circ}$ diameter) is $l_d \sim 10^{-12}$ times the brightness of 
sunlight. If 
sunlight is passed through an iron layer (or foil, respectively), then it  
is detectable with the naked eye only if it has passed not more than 
$$
( \ln (1/l_d ) + \ln ( \varepsilon_1 \varepsilon_2 \varepsilon_3 )) \delta 
\sim 170 \mbox{nm}. 
$$
Reflection losses by haze in the atmosphere further reduce this value. 

Kundt's observation of sunlight which penetrated through iron layers of 
up to 290nm thickness 
can hardly be explained by classical electrodynamics. 
Air bubbles within the metal layers cannot explain Kundt's observation, 
because air does not generate such a large rotation. Impurities, such 
as glass, which do generate an additional rotation, cannot completely be 
ruled out as the explanation. However, impurities are not a likely 
explanation, because Kundt was able to reproduce his observation by using 
several layers which he examined at various places. 

Quantum effects cannot explain the observation, because they decrease 
the penetration depth, whereas an increment would be required.

The observation may become understandable if Kundt has observed a 
second kind of electromagnetic radiation, the magnetic photon rays. 
I predict their penetration depth to be 
$$
\delta_m = \delta (c/v)^2 \sim 5\mbox{mm}.
$$
To learn whether Kundt has indeed observed magnetic photon rays, his 
experiment has to be repeated.

\subsection{The Experiment by Alipasha Vaziri}

\noindent
On February 22, 2002 between 15:30 and 16:30 local time 
of Vienna/Austria, Alipasha Vaziri tried an experiment 
to verify my predicted magnetic photon rays. 
As a light source he used a He-Ne laser of 1 milli Watt 
power and wavelength 632 nano meters. He coupled the 
light in a multi mode optical fibre with coupling 
efficiency of 70\%. The light came out at the other end. 
After 3 centi meters he coupled the light in a second 
multi mode glass fibre, also with coupling efficiency 
of 70\%. In front of the second optical fibre he 
placed an aluminium foil to shield the electric photon 
light. Behind the second optical fibre he placed an 
avalanche diode with 30\% efficiency for electric photon 
light of 632 nano meters wavelength as a detector. 

He did four sets of runs. Each run lasted for 10 seconds. 

In the first set the laser illuminated the foil.  The 
effective power of the laser was 56 micro Watts, because 
the sensitive area of the optical fibres was smaller than 
the cross-section of the laser beam. 
The counts of the 15 runs were: 
\begin{center}
350, 341, 339, 338, 337, 338, 331, 333, 336, 333, 
325, 327, 341, 335, 343.
\end{center}
For the second set the laser was off. The counts of the 
14 runs were: 
\begin{center}
344, 332, 329, 337, 332, 336, 338, 336, 343, 336, 
330, 344, 333, 338.
\end{center}
For the third set of experiments, he placed optical 
lenses between the two optical fibres to focus the 
laser beam. The effective power of the laser was 1 
milli Watt. The counts of these 17 foreground runs 
were:
\begin{center}
367, 343, 345, 356, 339, 348, 345, 355, 353, 358,  
346, 352, 345, 347, 342, 342, 345. 
\end{center}
For the fourth set, the optical lenses were placed  
between the optical fibres and the laser was off.  
The counts of the 15 runs were:  
\begin{center}
336, 337, 330, 345, 341, 345, 340, 337, 339, 343,
345, 337, 332, 340, 330.
\end{center}
In total, he made 44 background runs and 17 foreground 
runs. The mean background count rates were: 
\begin{center}
set 1: 33.65 counts/s

set 2: 33.63 counts/s

set 4: 33.85 counts/s

mean : 33.71 counts/s
\end{center}
The mean foreground count rate was: 
\begin{center}
set 3: 34.87 counts/s
\end{center}
Therefore the excessive count rate was 1.16 counts/s. 

The error bar can be estimated as follows. Two thirds 
of all data points should be within the one-sigma 
error bar, 95\% of all data points should be within 
the two-sigma error bar. The individual error bar 
is therefore 6 counts for the 44 background runs 
and 7 counts for the 17 foreground runs. The total 
error bar can be calculated by dividing the 
individual error bar through the square-root of the 
number of runs. Hence, the total error bar for the 
background is 0.9 counts, that of the foreground is 
1.7 counts. 

The count rates are therefore: 
\begin{center}
foreground : (34.87 $\pm$ 0.17) counts/s

background : (33.71 $\pm$ 0.09) counts/s

excess rate: ( 1.16 $\pm$ 0.19) counts/s
\end{center}
The statistical significance of the result is 
therefore 6 sigma. 

There is another interesting point. All of the 17 
foreground counts are larger than the mean of the 44 
background counts. The probability for this by pure 
chance is $1 : 2^{17} = 1 : 131072$. 

It is difficult to explain the small excess rate by 
conventional effects. 

(1) The statistical significance is 6 standard
    deviations. 

(2) The foreground runs were made between the second 
    and third background measurements. The mean count 
    rate of set 4, which directly followed the foreground 
    set, is close to those of sets 1 and 2. Therefore a 
    variability of the detector system (dark count rate) 
    is not a likely explanation. 

(3) Background set 4 was started directly after the 
    foreground set was terminated. The count rate dropped 
    simultaneously. Therefore it is unlikely that the 
    excessive count rate resulted from electronic noise 
    by equipment either inside or outside the laboratory. 

(4) The two optical lenses were used to focus the laser 
    beam, so they should have decreased effects of 
    stray light. It is therefore unlikely that the 
    excess is due to stray light. 

(5) The penetration depth of electric photon light of 
    632 nano meters in aluminium is only 3.68 nano meters. 
    Hence, the excess rate is not due to transmitted 
    electric photon light. 

(6) The excessive count rate is at least 7 orders of 
    magnitude too small to be explicable by electric 
    photon light which transmitted the aluminium foil 
    through a pinhole or hairline crack, respectively.
 
(7) Because of the second optical fibre, the electric 
    photon light of the laser cannot have heated the 
    avalanche detector.

\subsection{The Experiment by Roderic Lakes}

\noindent
The third experiment was performed by Roderic Lakes 
in Madison/Wisconsin in June 2002. As a light source 
he used a diode pumped YAG laser at 532 nano meters 
with 80 milli Watts of power. The detector was a 
photomultiplier with a quantum efficiency of 10\% for 
green electric photon light and a variable dark count 
rate between 5 and 30 counts/s. The diameter of the 
detector was 6.5 milli meters. An aluminium foil 
was placed directly in front of the detector. 

Roderic Lakes made 4 foreground sets and 3 background 
sets. Each set consisted of 6 runs. Each run lasted for 
10 seconds. The foreground and background sets 
alternated. 

The measured effect of the laser was 5 counts per second 
above background. 

It is difficult to explain this excess by conventional 
effects. 

(1) The foreground consisted of 5400 counts within 240 seconds. 
The mean foreground count rate was significantly greater than the 
mean background count rate. The background consisted of only 3200 
counts within 180 seconds.

(2) Foreground and background measurements alternated. 
    Therefore a variability of the detector is unlikely. 
    For the same reason, it is unlikely that the excess 
    results from noise of equipment either inside or 
    outside the laboratory.

(3) The penetration depth of electric photon light of 
    532 nano meters in aluminium is only 3.38 nano meters. 
    Hence, the excess rate is not due to transmitted 
    electric photon light.

(4) The excessive count rate is at least 8 orders of 
    magnitude too small to be explicable by electric 
    photon light which transmitted the aluminium foil 
    through a pinhole or hairline crack, respectively. 
    
I have to point out that neither Alipasha Vaziri nor 
Roderic Lakes claim to have detected a new effect. They 
wrote me that they disagree with my interpretation of their 
experiments (personal communications from Alipasha Vaziri 
and Roderic Lakes, June 12, 2003). Further experiments have to 
be done to ensure that the excessive count rates have 
indeed been generated by magnetic photon rays. 

\section{Consequences}

The observation of magnetic photon rays would be a multi-dimensional 
revolution in physics. Its implications would be far-reaching.

(1) The experiment would provide evidence of a second kind of electromagnetic 
radiation. The penetration depth of these magnetic photon rays is 
roughly one million times greater than that of
electric photon light of the same wavelength. Hence, these new rays may find 
applications in medicine where X-ray and ultrasonic diagnostics are 
not useful. X-ray examinations include a high risk of radiation damages, 
because the examination of teeth requires high intensities of 
X-rays and genitals are too sensible to radiation damages. Examinations 
of bones and the brain may also become possible.

(2) A positive result would provide evidence of an extension of (quantum) 
electrodynamics which includes a symmetrization of Maxwell's 
equations from 1873 [16].

(3) My model describes both an electric current and a magnetic current, 
even in experimental situations which do not include magnetic charges. 
This new magnetic current has a larger specific resistance in conductors 
than the electric current. It may find applications in electronics.

(4) The intensity of the magnetic photon rays should depend on 
the absolute velocity of the laboratory. The existence of the 
absolute velocity would violate Einstein's relativity principle of special 
relativity from 1905 [20]. It would be interesting to learn whether 
there exist further effects of absolute motion.

(5) The supposed non-existence of an absolute rest frame was the only 
argument against the existence of a luminiferous aether [20]. If the 
absolute velocity does exist, we have to ask whether aether 
exists and what its nature is.

(6) Magnetic photon rays may contribute to our understanding of 
several astrophysical and high energy particle physics phenomena 
where relativistic absolute velocities appear and where electric 
and magnetic photon rays are expected to be created in comparable 
intensities.

\nonumsection{Acknowledgements}

\noindent
I would like to thank Alipasha Vaziri (University of Vienna/Austria) 
and Roderic Lakes (University of Wisconsin/Madison) who tried the 
experiments to verify the magnetic photon rays.

\nonumsection{References}

\end{document}